\documentclass[twocolumn,english,floatfix,aps,prb]{revtex4}
\usepackage{graphicx}
\usepackage{amsmath}
\usepackage{amssymb}
\usepackage{color}
\makeatletter

\usepackage{babel}
\usepackage{algorithm}
\usepackage{algorithmic}

\newcommand\figref[1]{Fig.~\ref{#1}}
\newcommand\myeqref[1]{Eq.~(\ref{#1})}

\makeatother
\begin{document}

\title{Configuration Memory in Patchwork Dynamics for Low-dimensional Spin Glasses}

\author{Jie Yang, A. Alan Middleton}

\affiliation{Department of Physics, Syracuse University, Syracuse, New York 13244, USA}

\begin{abstract}
A patchwork method is used to study the dynamics of loss and recovery of an initial configuration in spin glass models in dimensions $d=1$ and $d=2$. 
The patchwork heuristic is used to accelerate the dynamics to investigate how models might reproduce the remarkable memory effects seen in experiment. 
Starting from a ground state configuration computed for one choice of nearest neighbor spin couplings, the sample is aged up to a given scale under new 
random couplings, leading to the partial erasure of the original ground state. The couplings are then restored to the original choice and patchwork
coarsening is again applied, in order to assess the recovery of the original state. Eventual recovery of the original ground state upon coarsening is 
seen in two-dimensional Ising spin glasses and one-dimensional clock models, while one-dimensional Ising spin systems neither lose nor gain overlap with 
the ground state during the recovery stage. The recovery for the two-dimensional Ising spin glasses suggests scaling relations that lead to a recovery 
length scale that grows as a power of the aging length scale.
\end{abstract}
\maketitle

Spin glass models are quintessential models for the study of disordered magnetic alloys, in the case where the interaction of magnetic moments has both 
frozen randomness and competition. The equilibrium and non-equilibrium behaviors of spin glass models are essentially different from those for homogeneous 
materials. For example, at zero temperature, the ground states do not exhibit conventional magnetic order but instead exhibit spatially randomly varying 
correlations of spins over long distances. The spin glass, whether it is a model or material, poses formidable challenges to the understanding of its 
complex non-equilibrium behaviors, including extremely slow dynamics and strongly  history-dependent effects.\cite{SGReviews} At low temperature, the 
full relaxation time for a spin glass material exceeds experimental time scales, though the slow approach to equilibrium can be studied in detail. A 
prototypical example of the resulting history-dependent effects is found in temperature cycling experiments.\cite{MemoryEx} Below the spin glass temperature, spin glass samples are always out of equilibrium. If the otherwise uniform-rate cooling of a spin glass material is paused at one temperature,
its magnetic susceptibility will slowly vary, ``aging" with time. Upon further cooling and subsequent reheating at a uniform rate, when the aging temperature is crossed, the magnetic susceptibility will fluctuate from a monotonic curve, moving towards its 
aged value. That is, the sample can ``recall'' its 
temperature history. Numerous explanations for this memory effect have been presented.\cite{Explanation1,Explanation2} Nonetheless, despite 
intensive study, many aspects of spin-glass dynamics remain poorly understood. As theoretical approaches, including droplet\cite{Droplet} 
and replica-symmetry-breaking\cite{SymmBreak} pictures, rely on distinct views of spin glasses and exact results are rare, numerical approaches have 
proven invaluable for verifying theories and motivating new ideas. \cite{SGAlgorithm}  

Generic simulation approaches for spin glass models can also be excessively slow, due to the computational complexity of spin glass models. Finding the 
exact ground states of spin glasses in the general case is an NP-hard problem and so appears to require exponential time to solve. Finding ground states 
in three dimension can be impractical for more than a few thousand spins.\cite{Barahona,ComComplexity,Liers} Direct Monte Carlo simulations, using 
single-spin-flip Glauber dynamics, have been used to explore the non-equilibrium spin configurations.\cite{PARISI} However, the evolution of a spin glass 
model under Glauber dynamics is extremely slow due to the high free-energy barriers between low free-energy configurations. Multiple applicable techniques 
for accelerating the process to equilibrium in spin glasses, such as parallel tempering \cite{Tempering} and simulated annealing,\cite{Annealing} have 
been developed but are also limited, though these techniques provide a wealth of information on the non-equilibrium behaviors of spin glasses. A fortunate 
exceptional case is the two-dimensional (2D) Ising spin glasses (ISG). Efficient algorithms that run in polynomial time do exist for exactly computing the 
equilibrium states of the 2DISG in the absence of a magnetic field.\cite{Barahona,MatchingCity} Here we exploit a fast algorithm for regional equilibration, ``patchwork dynamics",\cite{Patchwork} 
implemented here for zero temperature, to mimic the aging and memory effects in low-dimensional spin glass systems.

Patchwork dynamics is motivated by the expected dependence of free-energy barriers on length scale. For a domain with scale $\mathcal L$, the free-energy
barrier $\mathcal B$ for the reversal of all spins is anticipated to scale as ${\mathcal L}^\psi$, where $\psi$ is a barrier 
exponent.\cite{Droplet,EnergyBarrier} At temperature $T$, a domain will then survive for a characteristic time $t\sim e^{{\mathcal B}/T}$, as thermal 
excitation is required to conquer the free-energy barrier. Given this exponential dependence on $\mathcal L$, a strong separation of time scales is 
provided by an examination of geometrically separated length scales. Patchwork dynamics uses this separation to replace a dependence on time scales by 
a dependence on length scales. This approach equilibrates the system at a given length scale directly using efficient equilibrium algorithms, which 
overcome large energy barriers quickly, rather than waiting for local single-site dynamics to slowly reach that length scale. By equilibrating 
collections of spins on a succession of increasing length scales, patchwork dynamics approximates the coarsening process. It takes advantage of 
efficient equilibrium algorithms to investigate the non-equilibrium behaviors of spin glasses. The evolution of spin glasses configurations is sped up, 
in a heuristic fashion, by leaping over intermediate configurations with higher-free energies. We use this approach with the hope that it will provide us 
new insights into non-equilibrium behaviors of spin glasses.

In this paper, we investigate Ising spin glass systems on low-dimensional lattices: the square grid in two dimensions, the quasi-1D ``ladder", and the one-dimensional (1D) ``chain". We also examine the 1D $m$-state clock model spin glasses (CMSG). For the ISG, the Hamiltonian is 
$\mathcal {H_I}=-\sum_{\langle	ij\rangle}J_{ij}\sigma_i\sigma_j$; for the CMSG, the Hamiltonian has the form 
$\mathcal {H_C}=-\sum_{\langle	ij\rangle}J_{ij}\cos(\theta_j-\theta_i+\alpha_{ij})$. For all models, the couplings $J_{ij}$ between nearest neighbor spins 
$\langle ij\rangle$ are mean-zero independent Gaussian random variables. The $\sigma_i$ and $\theta_i$ are spin variables on a $d$-dimensional lattice with 
$\sigma_i=\pm1$ and $\theta_i=\frac{2\pi n}{m}$, where $n=1,\ldots,m$; the random noises $\alpha_{ij}$ between near neighbor spins for the CMSG are random 
variables chosen from a uniform distribution on the interval $[0,2\pi)$. Note that the ISG is equivalent to the standard CMSG with $m=2$ where the 
angle $\alpha_{ij}$ can be eliminated in the Hamiltonian. Highly efficient equilibrium algorithms exist for these spin glass systems in dimensions 
$d=1$ and $d=2$.\cite{DP,MatchingCity}

Patchwork dynamics can be employed to attempt to replicate aging and memory effects. These dramatic history-dependent effects are typically observed in 
temperature cycling experiments. Similar effects are expected to be seen when the couplings are perturbed and then reverted rather than perturbing the 
temperature.\cite{DisorderChaos1,DisorderChaos2} These behaviors are clearly related if temperature and disorder chaos\cite{BrayMoore} are related to 
the memory effects. At finite temperature, a small change in temperature leads to randomization of the coarse-grained couplings at the chaos scale. We 
mimic ``temperature cycling'' experiments by ``disorder cycling", replacing the temperature chaos by the disorder chaos at $T=0$.

Our simulation protocol is divided into two stages. The first stage is aging. We start with a ground state configuration for given couplings $J_{ij}$
as our initial spin configuration. In a finite sample, the ground states of the ISG with a continuous distribution for $J_{ij}$ are doubly degenerate (as accidental degeneracies have zero probability), distinguished by a global spin flip,\cite{flipSymmetry} 
while the $m$-state CMSG possesses $m$-times degenerate ground states, related by spin rotation symmetry. This finite degeneracy is a result of the continuous distribution of couplings that are used here, so that (with unit probability in the choice of random couplings) the energies of each configuration is distinct from configurations that are not related by global spin flips or rotations. The initial spin configuration is randomly 
chosen from the twofold ($m$-fold) ground state configurations (using the extended ground state with variable boundary conditions for the 
2DISG \cite{MatchingCity}). We then change the couplings $J_{ij}$ to an independently chosen set $J^{\prime}_{ij}$ , using disorder changes as a 
proxy for temperature changes. Patches are then applied using the new couplings $J^{\prime}_{ij}$ at a sequence of increasing scales $\ell$.
In each step of the sequence, we choose randomly located patches of length scale $\ell$, and update the spins in that patch to the lowest energy 
configuration, given fixed spins at the boundary of the patch. For each $\ell$, we apply $C(L/\ell)^d$ patches, with coverage $C=20$.
This ages towards a ground state for the new disorder $J'_{ij}$ that is uncorrelated with the initial ground state for bond couplings $J_{ij}$. The 
second stage after this aging stage is recovery. We halt the aging process at a chosen patch size, the aging scale, and then the couplings are reverted 
to $J_{ij}$. Recovery advances by optimizing patches on successive length scales $s=1,2,4,\ldots$, now using the original couplings. The maximum of the patch size $s$ 
used during recovery is $L/2$, for a system of size $L$.

\begin{figure}
\centering
\includegraphics[width=\columnwidth]{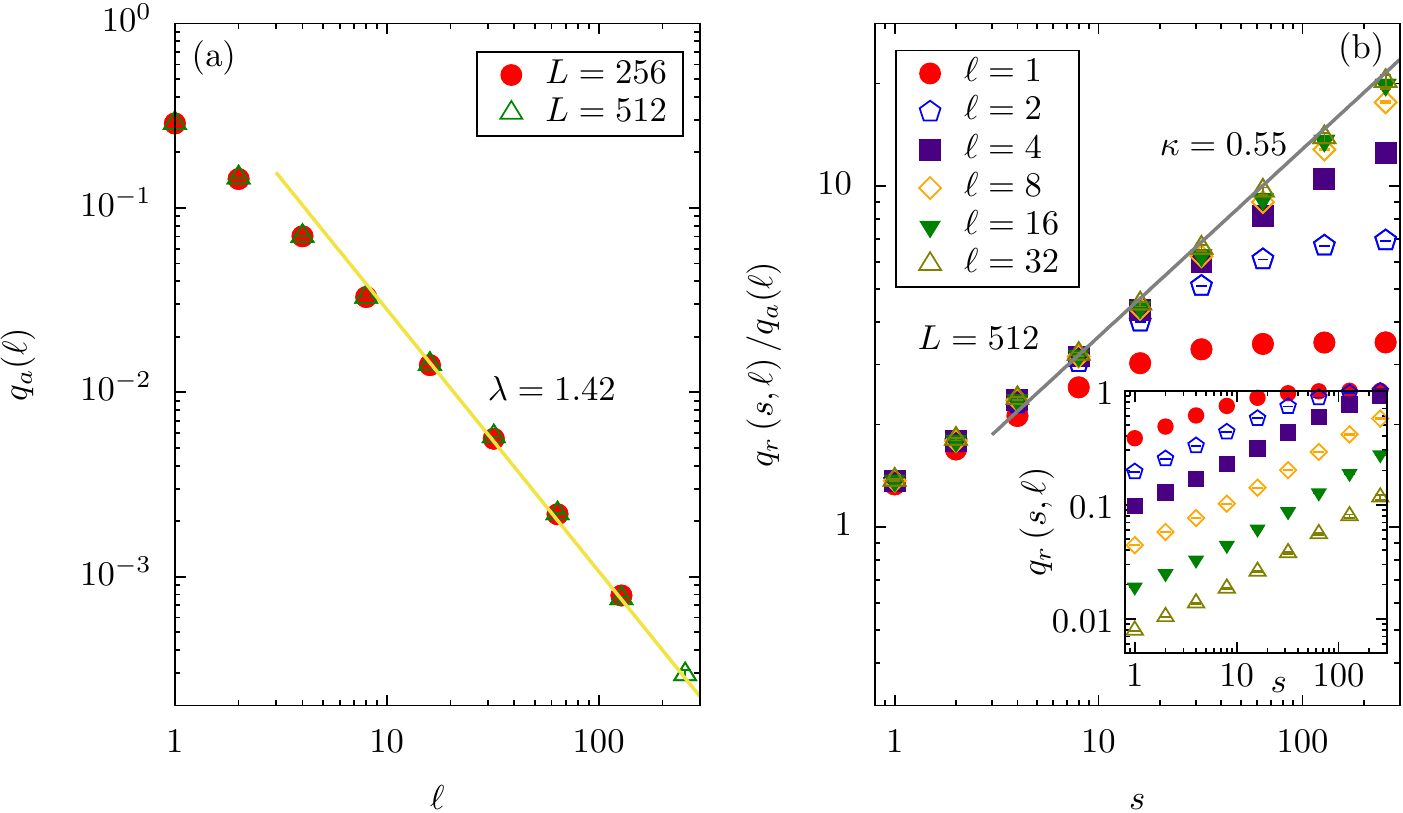}
\caption{(Color online).
Plot showing the sample-averaged spin overlap $q$ of 2D Ising spin glass configurations with the initial ground state in the aging and recovery stages, 
using numerical patchwork dynamics simulations as a heuristic for long time coarsening. (a) The spin configuration is initialized to a ground state configuration with couplings $J_{ij}$ and is then aged with uncorrelated couplings 
$J^{\prime}_{ij}$ with patch sizes $\ell$ geometrically increasing. The spin 
overlap $q_a(\ell)$ with the initial ground state state during aging decays consistently with a power law with a 
fitted exponent $-\lambda\simeq-1.42$. (b) For $L=512$, we cease aging stage at scales $\ell=1,2,4,8,16,32$, revert the couplings to 
$J_{ij}$, and then again apply patchwork dynamics. The spin configuration recovers to the original ground state as indicated by the increasing overlap, $q_r\left(s,\ell\right)$. This plot shows that the spin overlap $q_r\left(s,\ell\right)$ 
initially grows in a fashion consistent with $q_r\left(s,\ell\right)/q_a(\ell)\sim s^{\kappa}$ with best estimated $\kappa\simeq0.55$ at larger $\ell$ and $s$. The inset shows saturation of $q_r\left(s,\ell\right)\rightarrow1$ for smaller $\ell$ and large $s$.}
\label{fig:2Dq}
\end{figure}

\begin{figure}
\centering
\includegraphics[width=\columnwidth]{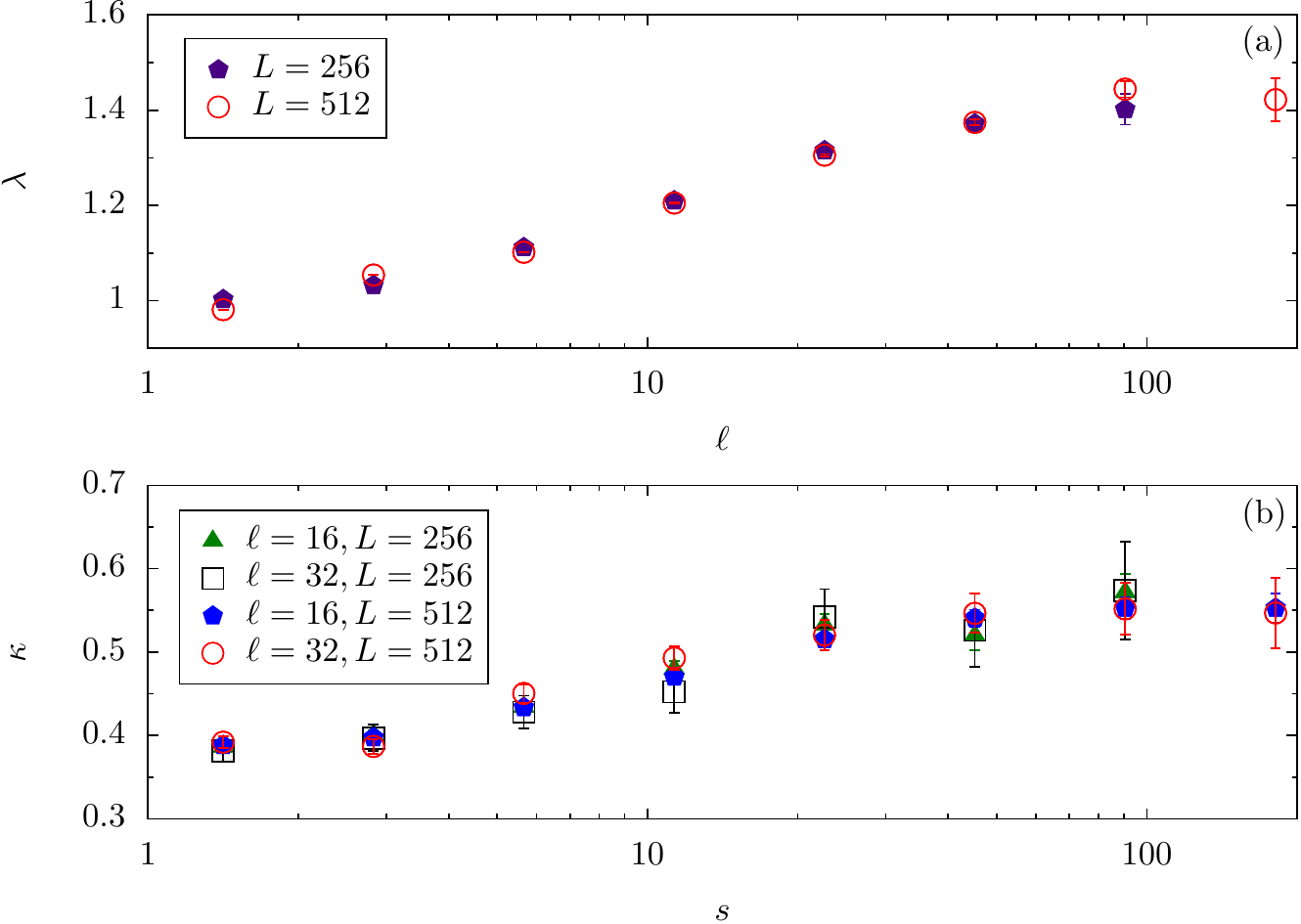}
\caption{(Color online). Plots of local exponents $\lambda$ and $\kappa$ derived from neighboring data points plotted in \figref{fig:2Dq} in the (a) 
aging and (b) recovery stages. The points are plotted versus effective patch scales $\ell$ and $s$, the geometrical mean of two subsequent patch scales used to determine the 
local exponents. The local exponents $\lambda(\ell)$ and $\kappa(s)$ appear to plateau for $\ell$ and $s$ greater $50$. The small $\ell$ and $s$ 
deviations are attributed to lattice effects. We extrapolate that under the limit of infinite sample sizes the exponents of 2D Ising spin glasses aging 
and recovery dynamics are $\lambda\simeq1.42\pm0.08, \kappa\simeq0.55\pm0.06$. Data shown are averaged over $10^5$ samples.
}
\label{fig:2Dslope}
\end{figure}

The sample averaged spin overlap $q=N^{-1}\overline{\sum_i \sigma^0_i \sigma_i}$ is a measure of the similarity of the initial ground state configuration $\sigma^0_i$ 
with subsequent configurations $\sigma_i$ for the ISG, where $N$ is the number of spins in the system. The overlap $q$ takes the maximal value $q=1$ for $\sigma_i=\sigma^0_i$, the minimal value $q=-1$ for $\sigma_i=-\sigma^0_i$, and has average value zero when $\sigma_i$ is uncorrelated with $\sigma^0_i$. The 2DISG has $N=L\times L$ spins arranged on a square 
lattice with toroidal boundary conditions \cite{MatchingCity}. The ladder IGS and the chain Ising spin system also have periodic boundary conditions and 
have $k\times L$ spins where $k$ is the number of layers of length $L$ with open boundaries in the layer direction ($k=2$ in the ladder ISG and $k=1$ in the chain system). 
In equilibrium, the overlap between two equilibrium configurations is used to extract thermodynamic behavior of spin glasses.\cite{overlapMiddleton} 
Here we use the overlap with the initial global ground state to quantify the aging and memory effects. For the 1DCMSG, the spin overlap is chosen to be 
$q=N^{-1}\overline{\sum_i \cos(\theta^0_i-\theta_i)}$. The $\theta^0_i$ and $\theta_i$ are spin variables in the initial state and subsequent states, 
respectively; the number of spins is $N=L$ and periodic boundary conditions are imposed. The notation $q_a(\ell)$ and $q_r\left(s,\ell\right)$ will be used to denote 
overlaps computed in the aging stage at scale $\ell$ and the recovery stage at scale $s$, respectively.

Our patchwork heuristic aging and recovery procedures were applied to the 2DISG with sample sizes $L=256, 512$. The behaviors of spin overlap $q_a(\ell)$ 
and $q_r\left(s,\ell\right)$ during aging and recovery shown in \figref{fig:2Dq}. In the aging stage, the decay of spin overlap $q_a(\ell)$ is well described by an asymptotic 
power law, $q_a(\ell)\sim {\ell}^{-\lambda}$. This result is consistent with the conjecture introduced by Fisher and Huse,\cite{Droplet} where $\lambda$ is 
an independent exponent of spin glass dynamics, and previous numerical results on 2D patch dynamics.\cite{Patchwork}

In the recovery stage, we expect that the spin configuration will approach a ground state for $J_{ij}$ as the patch size reaches $L$. However, it is 
undetermined whether the initial spin configuration $\sigma_i=\sigma^0_i$ or the reversed configuration $\sigma_i=-\sigma^0_i$ will be reached, given that the ground states 
are doubly degenerate. Since $q_r\left(s,\ell\right)>0$, our results indicate that samples approaching the initial spin configuration dominate those accessing the opposite 
ground state configuration, which statistically confirms that the 2DISG does have a memory of the initial spin configuration. We confirm that during the
recovery stage, until saturation, the growth of spin overlap is well described by another asymptotic power law, $q_r\left(s,\ell\right)/q_a(\ell)\sim s^{\kappa}$, until saturation to $q=1$ is approached.\cite{Patchwork}

To attain an improved estimate of the scaling exponents $\lambda$ and $\kappa$, we present the local exponents of spin overlap in \figref{fig:2Dslope}. 
These are the slopes, $\frac{\Delta ln (q_a(\ell))}{\Delta ln (\ell)}$ and $\frac{\Delta ln (q_r\left(s,\ell\right))}{\Delta ln (s)}$, between two neighbor patch sizes. The patch 
sizes $\ell$ and $s$ on the horizontal axes in \figref{fig:2Dslope} are the geometrical mean of the two subsequent patch sizes used to calculate the local 
exponent. Given our high precision and our use of local slopes (rather than a simpler power law fit), we can see that $\lambda$ and $\kappa$ are not constant at small $\ell$ and $s$ and appear to crossover to a large distance limit for 
$\ell,s\geq50$. The estimated values are 
$\lambda=1.42\pm 0.08,\kappa=0.55\pm 0.06$, where systematic uncertainties are subjective and exceed the statistical error estimates at these length scales. We use these values to show consistency with power law fits to the largest scale points in \figref{fig:2Dq}.

We observe in these latest simulations that as the maximum patch size $\ell$ decreases, the spin overlap in the recovery stage $q_r\left(s,\ell\right)$ diverges from the 
asymptotic power law seen for large $\ell$. Instead, the spin overlap $q_r\left(s,\ell\right)$ crosses over to a plateau as patch sizes $s$ increase at fixed small values of 
$\ell$. The existence of this plateau is consistent with the spin configurations reaching the global ground state, which implies that the spin overlap 
$q_r\left(s,\ell\right)\rightarrow1$ at sufficiently large $s$. This suggests a scaling form for recovery that incorporates an initial power law increase followed by a plateau 
at $q_r\left(s,\ell\right)=1$. We emphasize that this scaling form assumes full recovery to the original spin direction with probability near 1, consistent with simulations 
for $s\gg \ell$. We also assume that the power law fits for aging and recovery are valid, so that the starting spin overlap $q_a(\ell)$ at the start of the 
recovery stage is $\ell^{-\lambda}$. This overlap then grows $q_r\left(s,\ell\right) \sim s^{\kappa}{\ell}^{-\lambda}$ during recovery until this overlap $q_r\left(s,\ell\right)$ approaches unity.
This leads to a crossover recovery scale $s_c$ at patch size $s_c\sim {\ell}^{\frac{\lambda}{\kappa}}$ at which the spin overlap $q_r\left(s,\ell\right)$ approaches $1$. The 
existence and scaling of this recovery scale is a central result of this current work. As $\lambda>\kappa$, the coarsening length scale for recovery $s_c$ 
grows faster than the aging scale $\ell$. The aging process erases the initial state sufficiently that the recovery scale, defined as the point where the 
overlap exceeds some fixed fraction of unity, say $1/2$, must span a scale many times that of the aging scale, in the 2DISG.

\begin{figure}
\centering
\includegraphics[width=\columnwidth]{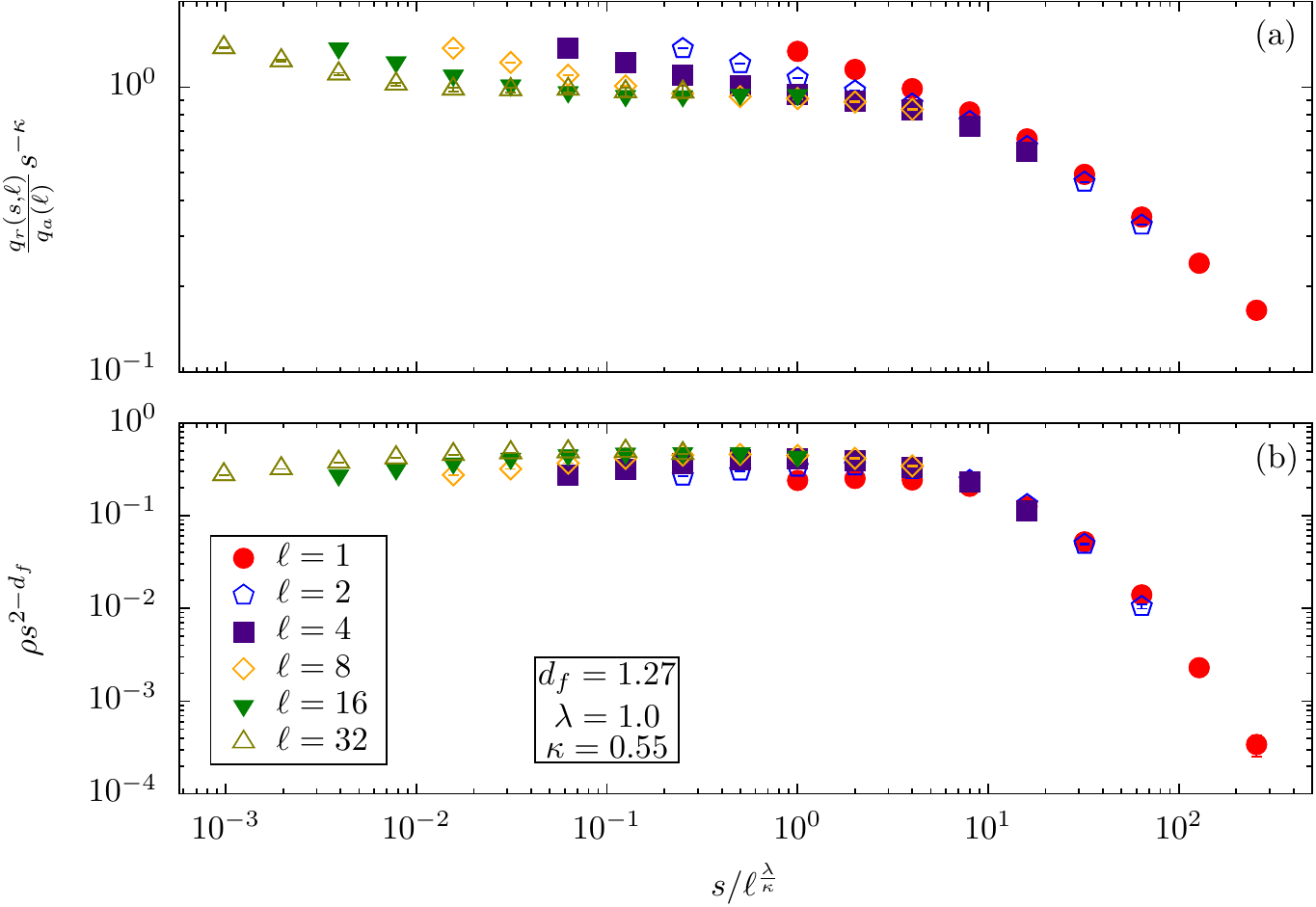}
\caption{(Color online).
Scaling collapses for (a) the spin overlap $q_r\left(s,\ell\right)$ and (b) domain wall bond density $\rho$ in the recovery stage for sample size $L=512$. For the patch size 
$s$ smaller than the crossover length $s_c$, the spin overlap scales as $\frac{q_r\left(s,\ell\right)}{q_a(\ell)}\sim s^{\kappa}$ with $\kappa\simeq0.55$ and the domain wall 
bond density scales as $\rho\sim s^{d_f-2}$ with $d_f\simeq1.27$. The statistics $\frac{q_r\left(s,\ell\right)}{q_a(\ell)}s^{-\kappa}$ and $\rho s^{2-d_f}$ are used to check the 
scaling $s_c\sim{\ell}^{\frac{\lambda}{\kappa}}$. The deviations at small scaling parameters are ascribed to lattice effects. We adopt here 
$\lambda\simeq1.0$, which is the local exponent for small aging patch size $\ell$, where we see a divergence from a simple power law at larger $\ell$, 
where $\lambda$ approaches $\lambda\approx1.4$. The two statistics exhibit scaling collapses at an identical significant patch size $s_c$, which is 
consistent with our expectation.
}
\label{fig:2D_collap}
\end{figure}

Assuming a single recovery scale, scaling forms for both the spin overlap $q_r\left(s,\ell\right)$ and the domain wall bond density $\rho$ in the recovery stage are fixed.
Scaling collapses for these statistics are shown in \figref{fig:2D_collap}, based on this description. The domain wall density is the fraction of bonds 
belonging to domain walls that separate ground states $A$, with spins $\sigma^0_i$, and $\bar{A}$, with spins $-\sigma^0_i$. The number of domain wall 
bonds in the 2DISG are taken to scale as $s^{d_f}$, with the domain walls for domains at recovery scale $s<s_c$ are taken to have the standard 2DISG 
domain wall fractal dimension $d_f\simeq1.27$.\cite{FractalDomain1,FractalDomain2} The domain wall bond density $\rho$ thus scales as $s^{d_f-2}$ until 
the recovery scale $s_c$ is reached. To better describe our data following scaling collapses, we use the estimate $\lambda=1.0$, the value of the 
effective exponent for small $\ell$, given that our data show the plateau in \figref{fig:2Dq} for small aging patch sizes, such as $\ell=1,2,4$, along 
with the large $s$ value $\kappa=0.55$. We do not observe the scaling collapse for larger $\ell$, because under that circumstance the patch size $s_c$ that
scales $s_c\sim {\ell}^{\frac{\lambda}{\kappa}}$ exceeds $L/2$ and thus is unreachable for our system sizes. The spin overlap $q_r\left(s,\ell\right)$ and domain wall bond 
density $\rho$ in \figref{fig:2D_collap} both indicate an identical patch scale $s_c$ where $q_r\left(s,\ell\right)$ approaches $1$ and $\rho$ vanishes rapidly from the 
scaling collapses, which corroborates our anticipated connection. Those points slightly deviating from a single curve for the scaling collapse at fixed values 
of $s$ smaller than $s_c$ are believed to stem from lattice effects. Overall, this scaling collapse with effective exponents for the domain wall density 
support a scaling picture for the recovery scale and for the relationship between domain wall density and recovery scale.

We have also carried out simulations to explore how the recovery depends on the sample preparation. For the results above (and those below for lower dimensional 
systems), we studied recovery after a specific coarsening process. This process decreases the spin overlap with the initial configuration by growing domains 
at the patch scale that are given by an independent set of spin couplings. For comparison, we have prepared samples by randomly selecting a subset of the initial 
spins and then flipping all of the spins in that subset. The size $F$ of the subset chosen determines the overlap $q$ with the initial configuration, via the 
relation $q=1-2F/N$. There is no spatial correlation in which spins are flipped or remain unchanged in this process (by contrast, such a correlation could conceivably result from patchwork aging). The overlap can be tuned arbitrarily without going through 
an aging process. We find that the recovery $q_r(s)$ from the chosen values for $q$ are described by the same exponent $\kappa$, i.e., $q_r(s)\sim q s^\kappa$. This 
suggests that the memory left after the aging process is not specific to coarse-graining towards another ground state. Isolated, uncorrelated remnant spins can 
be used to recover the initial configuration with the same quantitative growth of overlap.

To explore the effect of dimensionality on these non-equilibrium behaviors of spin glasses, we investigate memory effects in Ising spin systems with lower 
dimensionality, including the frustrated ladder ISG and the chain Ising spin system.
Patchwork dynamics drives the spin configuration to a state that is low energy at the given scale for the given couplings. In the limit of coverage $C\rightarrow \infty$, the configuration is minimal with respect to arbitrary changes of spins up to that scale. This scale-dependent optimization results in domain walls separating subsets of
global ground states. This is consistent with the droplet picture used by Fisher and Huse,\cite{Droplet} where the energy and geometry of the domain walls depends on dimensionality, and the droplet energy and geometry affects equilibration and aging.\cite{Droplet}
These ladder and chain systems are both one-dimensional at large scales. Their ground states both consist of spins with statistically random orientations, as in the 2DISG. Nonetheless, there is 
an essential difference between their ground states. In the ground states of the chain Ising spin system, all bonds but for at most one are satisfied. Therefore,
the chain Ising spin system is not a spin glass, as there is no frustration. A bond is 
satisfied when neighbor spins on that bond are parallel if the bond coupling $J_{ij}$ is positive, or they are in opposite orientations if $J_{ij}$ 
is negative. Frustration results only from boundary conditions. The situation where a single bond with the minimum absolute value of $J_{ij}$ is 
unsatisfied in the ground state of the chain system occurs with probability $\frac{1}{2}$ due to the randomness of  the $J_{ij}$. With respect to the 
ladder ISG, conversely, frustrated plaquettes\cite{Barahona} commonly exist in its ground states. Bonds belonging to a frustrated plaquette cannot be 
all satisfied. Thus, the ground states of the ladder ISG include a finite density of unsatisfied bonds.
Though the chain Ising system does not have frustration, the effects of disorder remain important and, as we will see, the recovery of
memory in the chain is quite similar to that in the ladder and can be analyzed.  

\begin{figure}
\centering
\includegraphics[width=\columnwidth]{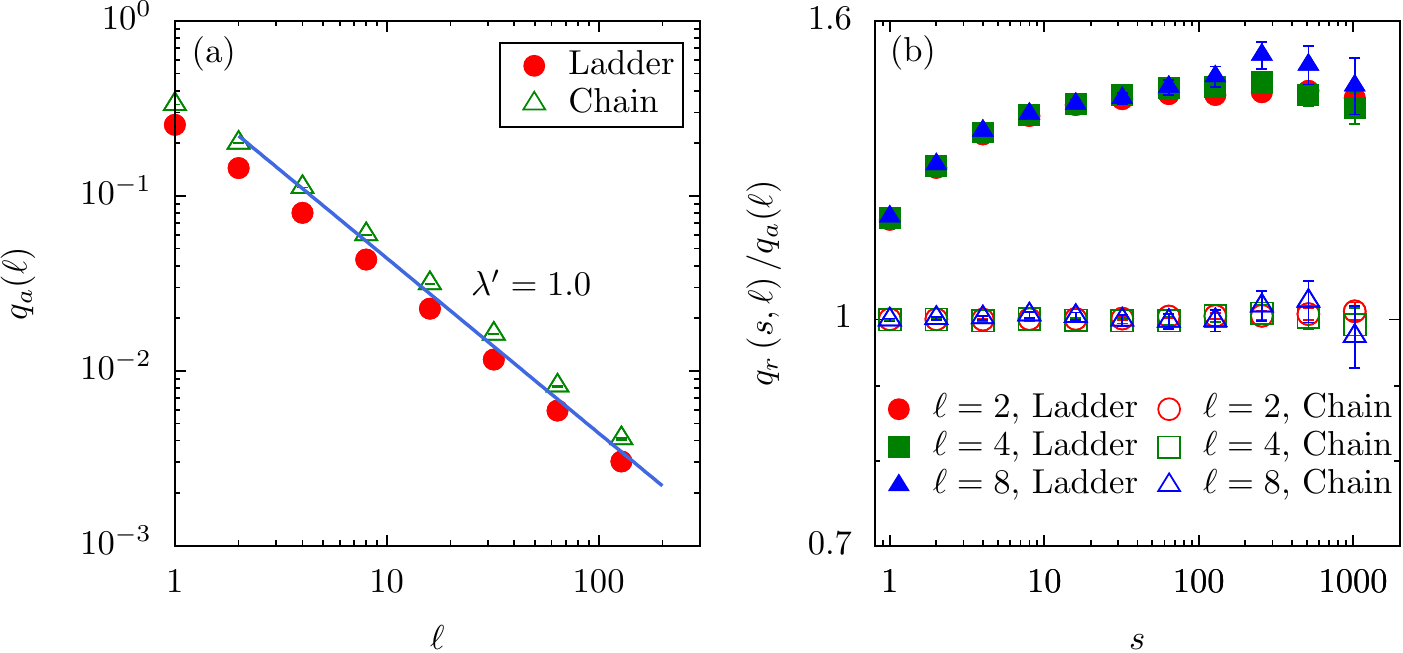}
\caption{(Color online).
Patchwork dynamics simulations on the ladder Ising spin glass and the chain Ising spin system for sample size $L=2^{17}=262\,144$. (a) In the aging stage spin overlap 
$q_a(\ell)$ of the $k=2$ ladder and the $k=1$ chain behave similarly and are well interpreted by a power law, which is $q_a(\ell)\sim {\ell}^{-\lambda^{\prime}}$ with
$\lambda^{\prime}\simeq1.0$. (b) In the recovery stage spin overlap $q_r\left(s,\ell\right)$ of the ladder and chain have somewhat distinct behaviors. In both cases, 
the aging stage is terminated at various scales $\ell=2,4,8$. The ladder ISG exhibits a weak initial memory effect. The statistic 
$q_r\left(s,\ell\right)/q_a(\ell)$ slightly grows and crosses over to a plateau of $1.5$, as is seen in the upper curves. For the chain system, the statistic $q_r\left(s,\ell\right)/q_a(\ell)$ stays constant at $q_r\left(s,\ell\right)/q_a(\ell)\simeq1$, which 
indicates that the chain system does not possess memory recovery.  
}
\label{fig:1D_q}
\end{figure}

We applied the patch aging and recovery protocols to the two-layer ($k=2$) ladder ISG system and to the chain Ising spin system, for sample sizes $L\leq2^{17} = 262\,144$. 
The evolutions of spin overlaps for aging and recovery are displayed in \figref{fig:1D_q}. The spin overlap $q_a(\ell)$ for both the ladder and the chain 
in the aging stage appear to follow power law decays, as seen in the 2DISG. The power laws for the two one-dimensional cases are similar to each other: 
The spin overlap $q_a(\ell)$ for both ladder and chain systems are both well described by a single power law, $q_a(\ell)\sim{\ell}^{-\lambda^{\prime}}$ with a  best 
fitted exponent $\lambda^{\prime}=1.00\pm0.03$. However, in the recovery stage the spin overlap $q_r\left(s,\ell\right)$ behave distinctly from each other at small recovery 
scales $s$. The recovery also differs at larger scales from that seen for the 2DISG (at all scales). To study the recovery of the spin overlap, we interrupt 
the aging at $\ell=2,4,8$ and then revert the couplings to the original $J_{ij}$.
The recovery statistic $q_r\left(s,\ell\right)/q_a(\ell)$ of the ladder ISG grows slightly and crosses over to 
a plateau $q_r\left(s,\ell\right)/q_a(\ell)\simeq1.5$.
Compared to the 2DISG, the memory effects for the ladder ISG are weak and apparently do not include a power law growth at large recovery patch 
sizes, only a plateau value for the recovery ratio $q_r/q_a$ that is independent of $L$.
We have also studied the ISG with more layers, such as $k=3$, where memory effects are apparently stronger than those of the two-layer ladder ISG, 
but we still find a plateau to a fixed recovery ratio that increases with $k$. It is plausible to infer that the capability of memory grows and trends towards the 
full recovery seen in the 2DISG, as $k$ increases.
The $k=1$ chain shows no initial recovery. The statistic $q_r\left(s,\ell\right)/q_a(\ell)$, the ratio of overlap in recovery to the overlap 
at maximal aging, of the chain system does not grow but stays fixed at $1$ for all length scales $s$, within our numerical error. This implies that the chain system does not show either 
any memory effects or any further aging under recovery coarsening. 

\begin{figure}
\centering
\includegraphics[width=\columnwidth]{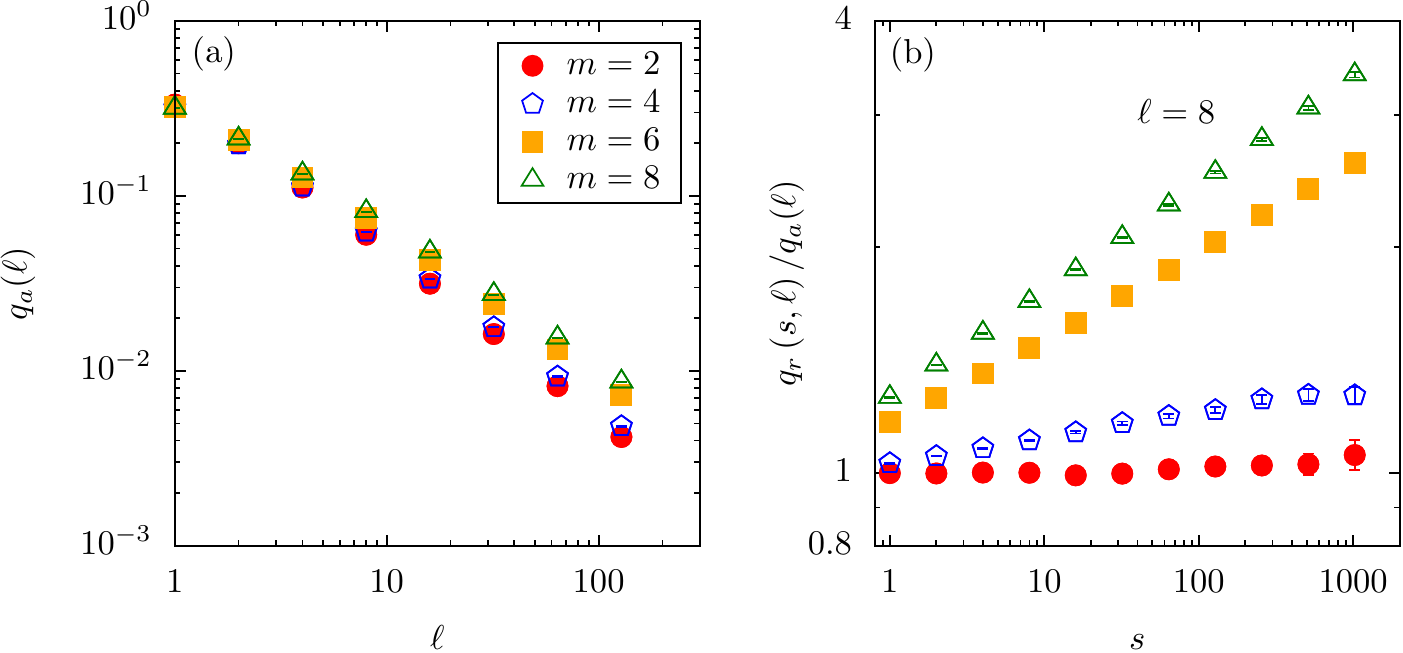}
\caption{(Color online).
Spin overlap $q$ of 1D $m$-state clock model spin glasses in patchwork dynamics. The sample size is $L=2^{17}=262\,144$ and the number of spin states 
is $m=2,4,6,8$. (a) The spin overlap $q_a(\ell)$ decay as power laws $q_a(\ell)\sim {\ell}^{-{\lambda}_{\mathcal C}}$ in the aging stage. The exponent 
${\lambda}_{\mathcal C}$ slightly decreases, as $m$ increases. (b) The spin overlap $q_r\left(s,\ell\right)$ in the recovery stage show significantly difference for 
various $m$. In the data displayed, we cease the aging at $\ell=8$. The absence of memory effect is indicated for $m=2$ as in the case of the chain 
Ising spin glasses. The spin overlap $q_r\left(s,\ell\right)$ for $m=4,6,8$ can be well described by power laws $q_r\left(s,\ell\right)/q_a(\ell)\sim s^{{\kappa}_{\mathcal C}}$, with small 
$\kappa_{\mathcal C}$. The 1D clock model spin glasses show the recovery of memory for $m=4,6,8$. The exponent ${\kappa}_{\mathcal C}$ grows as the
spin degrees of freedom $m$ increase (see text), which indicates weak reinforcement of memory with the increase of $m$.
}
\label{fig:cmsg_q}
\end{figure}

We find it interesting that the chain Ising spin system shows no memory in our simulations. To explain this absence of memory, with the goal of possibly explaining
the plateau in the $k>1$ one-dimensional systems, we calculate analytically the expected 
change of spin overlap when patches of arbitrary size $s$ are applied during the recovery stage. This expected change can be computed for an arbitrary 
spin configuration. Given a spin configuration in a given sample with fixed $J_{ij}$, we define $\Delta_i$ as the change of spin overlap $q_r\left(s,\ell\right)$ caused 
by optimizing the spins in a patch with size $s$ starting at position $i+1$, that is, with boundary spins $\sigma_i$ and $\sigma_{i+s+1}$ fixed. The average 
$\langle\Delta_i\rangle$ is then defined as the expectation value of the change in the spin overlap, where the angle brackets indicate an average over 
the position $i$. We will use the symbol $+$ for a spin of the same sign as in the chosen ground state configuration; the symbol $-$ means a spin has the 
opposite orientation. The overlap can be found from the sum of the number $N_+$ of $+$ spins, with $q=L^{-1}(2N_+ -1)$. There are $L$ different patches 
with size $s$ and they can be classified into four categories, one for each patch boundary condition. Boundary conditions for these 
categories are indicated by $++$, $--$, $-+$ and $+-$, with the two symbols referring to the relative orientation of two boundary spins for the linear 
patch with respect to the ground state configuration. The number of four different categories of patches in a given sample and given $s$ are respectively 
$n_{++}$, $n_{--}$, $n_{-+}$ and $n_{+-}$. When a patch is applied, the value of $\Delta_i$ depends on the category and the spins inside the patch.
All spins in patches with boundary conditions $++$ will be set to $+$; spins in patches with boundary conditions $--$ will have a post-patch orientation 
of $-$; for patches with boundary conditions $+-$ or $-+$, a domain wall will result at $b(i,s)$ where $b(i,s)$ is the relative location for the local 
minimum of $J_{ij}$ within the given patch. The sign of the spin with respect to the ground state changes between the spin with index $i+b(i,s)$ and the 
spin with index $i+b(i,s)+1$. We use the notation $\sum_{i,+-}$ to indicate a sum over all $i$ restricted to patches of type $+-$, for example.  When a patch 
forces a spin $\sigma_k$ to be $+$, the change in overlap is $(1-\sigma_k)/L$. and when a patch forces a spin $\sigma_k$ to be $-$, the change in overlap is $(-1-\sigma_k)/L$. 
Thus, given a uniform probability over location for the placement of patches, $\langle\Delta_i\rangle$ can be written as
\begin{align}
\langle\Delta\rangle
&= \frac{1}{L} \sum \Delta_i \nonumber \\
&= \frac{1}{L} \Biggl(\sum_{i,++}\sum_{k=i+1}^{i+s}(1-\sigma_k) + \sum_{i,--}\sum_{k=i+1}^{i+s}(-1-\sigma_k) \nonumber \\ 
&\: + \sum_{i,-+} \left[\sum_{k=i+1}^{i+b(i,s)}(-1-\sigma_k) + \sum_{k=i+b(i,s)+1}^{i+s}(1-\sigma_k)\right] \nonumber \\
&\: + \sum_{i,+-} \left[\sum_{k=i+1}^{i+b(i,s)}(1-\sigma_k) + \sum_{k=i+b(i,s)+1}^{i+s}(-1-\sigma_k)\right]\Biggr) \nonumber \\
&= \frac{s}{L} \left(-\sum \sigma_k + n_{++} - n_{--} + n_{-+} - n_{+-}\right) \nonumber \\
&\: +\frac{2}{L} \left(\sum_{i,+-}b(i,s)-\sum_{i,-+}b(i,s)\right) \,.
\label{eq:delta_q1}
\end{align}
Letting $N_-$ be the number of $-$ spins, the number of patch categories is constrained by the equations
\begin{align}
2n_{++}+n_{+-}+n_{-+} &= 2N_+\label{constraint1}\\
2n_{--}+n_{+-}+n_{-+} &= 2N_- \label{constraint2} \\
N_+-N_- &=\sum \sigma_k \,,\label{constraint3} 
\end{align}
from counting the contributions of each type of patch category to the total number of up or down spins (\myeqref{constraint1} and \myeqref{constraint2}) 
and from the definitions of $N_+$ and $N_-$ (\myeqref{constraint3}). It follows from Eqs.\ (\ref{constraint1}-\ref{constraint3}) that
\begin{align}
-\sum \sigma_k + n_{++} - n_{--} &= 0 \,.
\label{eq:eq1}
\end{align}
It is also true, for a given sample, due to the periodic boundary condition, that
\begin{align}
n_{-+} &= n_{+-} \,.
\label{eq:eq2}
\end{align}
To show this, the set of sites $\{1,\ldots,L\}$ can be decomposed into distinct orbits $O_1, O_2, \ldots$, where each orbit $O_m$ is a sequence of sites
$O_m=\{i_m+k(s+1)\mod L| k=0,1,\ldots\}$. These orbits are sets of the left ends of patches where the patches share boundary points. As $L$ is finite, 
each $O_i$ is finite. In each orbit $O_i$, there must be an equal number of $+-$ and $-+$ patches. Since $n_{+-}$ and $n_{-+}$ is the sum of these counts 
over the set of orbits, we have $n_{-+}=n_{+-}$.

Using \myeqref{eq:eq1} and \myeqref{eq:eq2}, we can write the expected change in overlap for a patch of size $s$ in a given sample, 
$\langle\Delta(J_{ij})\rangle$, as 
\begin{align}
\langle\Delta(J_{ij})\rangle &= \frac{2}{L} \left(\sum_{i,+-}b(i,s)-\sum_{i,-+}b(i,s)\right) \,.
\end{align}
The distance to the domain wall for mixed boundaries $b(i,s)$ can take any integer value in the range $[0,L-1]$. In an individual sample, it is likely 
that $\sum_{i,+-}b(i,s)\neq\sum_{i,-+}b(i,s)$. However, if we average over a large number of samples, the mean change of the spin overlap 
$\langle\Delta(J_{ij})\rangle=0$, due to the statistical left-right symmetry that does not on average distinguish $-+$ from $+-$ boundary points. We 
conclude that during patch dynamics for the 1D Ising spin system, the overlap with the ground state for the $J_{ij}$ used does not change. This analytic result is 
consistent with our simulation results for the recovery phase.

In the ladder system, there is an initial recovery seen in $q_r/q_\ell$, but this ratio reaches a fixed value at larger recovery patch size $s$. This plateau is plausibly explained using our observed and proven fixed value for $q_r/q_\ell$ for the 1D chain system, through universality. The effective dimensionality of domain walls is zero in both the ladder and chain systems and the cost of domain walls is finite in both cases. This suggests that an initial recovery in the ladder system occurs for patch sizes $s$ not much larger than the plaquette size, the scale where frustration is evident and the spins are behaving roughly as in the two-dimensional case. At large $s$, the frustration of the spin plaquettes is not relevant, as domain walls have finite cost and are of transverse size less than the patch scale. The recovery then approaches that for the 1D chain, with little additional recovery of the initial state.

We have also explored how the symmetry of the spin degrees of freedom (DOF) affects the recovery of memory in spin glasses. We employ our patch 
aging/recovery protocols on the 1DCMSG with varying number of single spin states, $m=2,4,6,8$. In the ground states of the 1DCMSG with $L>2m$, the bonds 
are mostly all satisfied, as the periodic boundary conditions can lead up to $m$ unsatisfied bonds. These ground states are $m$-fold degenerate. The spin 
overlap in coupling cycling simluations are exhibited in \figref{fig:cmsg_q}. In the aging stage, the spin overlap $q_a(\ell)$ is well fitted by power laws, 
$q_a(\ell)\sim{\ell}^{-{\lambda}_{\mathcal C}}$. A local exponent analysis shows that, as the number of spin states $m$ increases, the local power law exponent 
${\lambda}_{\mathcal C}$ marginally decreases with increasing $m$, though perhaps within systematic uncertainties. For example, quantitatively, the fitted 
exponent ${\lambda}_{\mathcal C}\simeq0.97, 0.94$ for $m=2,4$ respectively. We then study the behavior of spin overlap $q_r\left(s,\ell\right)$ in the recovery stage after 
halting the aging at $\ell=8$. The 1DCMSG with $m=2$ exhibits no memory effects, as in the case of the chain Ising spin system, though the Hamiltonian is slightly different 
from the chain case. For $m>2$, the 1DCMSG exhibits memory effects and the statistic $q_r\left(s,\ell\right)/q_a(\ell)$ is plausibly described by a small power law growth for recovery, 
$q_r\left(s,\ell\right)/q_a(\ell)\sim s^{\kappa_{\mathcal	C}}$. A logarithmic growth of the recovery ratio with $r$ is possible, though the fit is less good. The strength of memory 
exhibits some enhancement as $m$ increases over the range we studied , as is seen in \figref{fig:cmsg_q}; the estimated scaling exponent 
${\kappa}_{\mathcal C}\simeq0.14$ with $m=8$ is greater than ${\kappa}_{\mathcal C}\simeq0.12$ with $m=6$.

In conclusion, patchwork dynamics provides an efficient heuristic technique to investigate the non-equilibrium behaviors of large spin glass systems, which 
are more difficult to study at large scales using conventional Monte Carlo methods. We use patchwork dynamics to mimic the cooling and reheating 
processes on spin glass materials, by varying the bond couplings in a random fashion. The spin overlap and domain wall density show memory effects, as magnetic susceptibility does 
in experiments.\cite{MemoryEx} We expand upon the scaling description of the aging and recovery stages, consistent with numerical evidence. In particular, 
we identify a recovery scale $s_c$ which grows as a power law of the aging scale $\ell$, $s_c\sim \ell^{\lambda/\kappa}$, where $\lambda$ is the decay exponent for 
the spin overlap during aging and $\kappa$ is the overlap recovery exponent. As the exponent ratio $\lambda/\kappa \approx 2.58$, the recovery scale grows much 
more quickly than the aging scale. Thus recovery of the original spin direction requires coarsening to a scale that grows faster than the aging scale that 
causes erasure. We note that the recovery exponent $\kappa$ is independent of whether the erasure is caused by coarsening through aging or through random 
flipping of the spins. We have also studied how the dimensionality and the spin symmetry affect the strength of memory in various spin glass models. It is 
notable that the chain system shows the exact absence of memory effects. The ladder ISG, effectively one-dimensional at long scales but with local frustration, exhibits a plateau in recovery, i.e., the memory effect is only a finite multiple of the aged memory, so that full recovery does not occur. Simulations results from the one-dimensional clock model spin glass  indicate that memory effects become stronger as the number of spin states increases.

\textit{Acknowledgments}. This work was supported in part by the National Science Foundation Grant No. DMR-140937. This work was carried out primarily using the Syracuse University campus OrangedGrid , a computing resource supported by NSF award No. ACI-1341006.

\end{document}